\documentclass[10pt,twocolumn,letterpaper]{article}

\usepackage{cvpr}              % To produce the CAMERA-READY version
\usepackage[accsupp]{axessibility}
\usepackage{graphicx}
\usepackage{amsmath}
\usepackage{amssymb}
\usepackage{booktabs}
\usepackage{multirow}
\usepackage{adjustbox}
\usepackage{wrapfig}
\usepackage{diagbox}
\usepackage{epsfig}
\usepackage{extpfeil}
\usepackage{makecell}
\usepackage{colortbl}
\usepackage{float}
\usepackage{wrapfig}
\usepackage{booktabs}
\usepackage{tabulary,overpic,xcolor}

\definecolor{cvprblue}{rgb}{0.21,0.49,0.74}
\usepackage[pagebackref,breaklinks=true,colorlinks,bookmarks=false,citecolor=cvprblue,linkcolor=cvprblue]{hyperref}

\usepackage[capitalize]{cleveref}
\usepackage{algorithm}
\usepackage{algorithmic}

\newcommand{\ul}[1]{\underline{\smash{#1}}}
\definecolor{color3}{rgb}{0.95,0.95,0.95}
\definecolor{color4}{rgb}{0.96,0.96,0.86}
\definecolor{color5}{rgb}{0.90,0.90,0.90}

\crefname{section}{Sec.}{Secs.}
\Crefname{section}{Section}{Sections}
\Crefname{table}{Table}{Tables}
\crefname{table}{Tab.}{Tabs.}

\def\confName{CVPR}
\def\confYear{2024}

\title{Structure-Aware Sparse-View X-ray 3D Reconstruction}
\author{Yuanhao Cai,~~ Jiahao Wang,~~ Alan Yuille$^{\dag,*}$, Zongwei Zhou$^*$,~~ Angtian Wang \\
	Johns Hopkins University
}

\begin{document}
	
\twocolumn[{
	\maketitle
	\centerline{\hspace{-2mm}
		\includegraphics[width=1\linewidth,trim={4pt 4pt 4pt 4pt}]{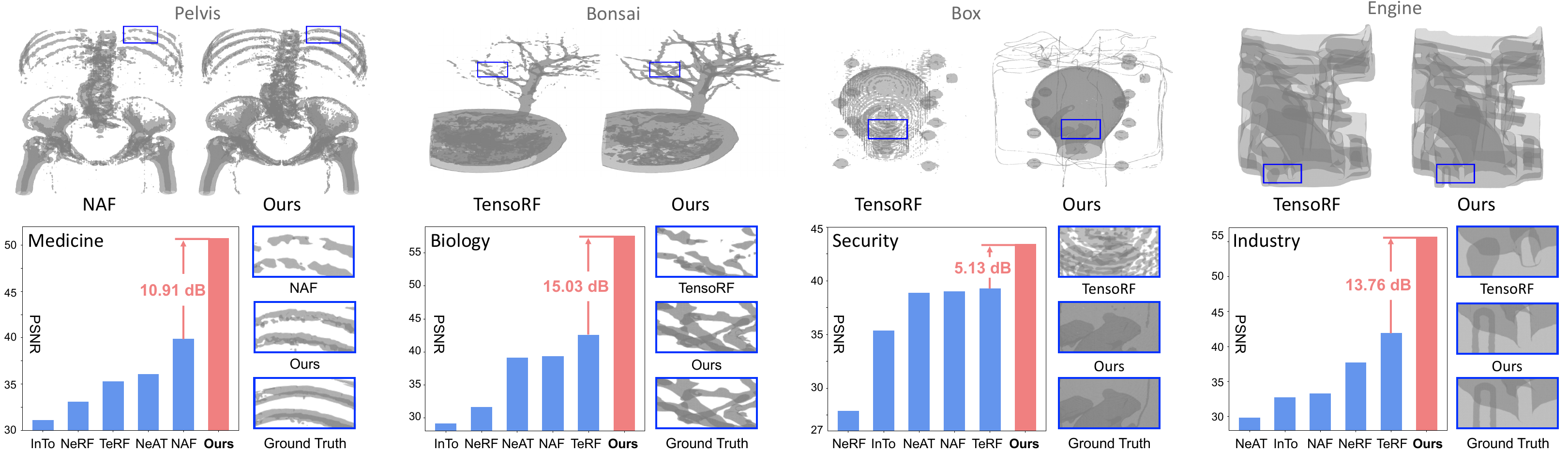}}
	\vspace{-1mm}
	\captionof{figure}
	{
		\small Comparisons of X-ray novel view synthesis. On the collected X3D dataset, our method surpasses state-of-the-art algorithms including InTo (InTomo~\cite{intratomo}), NeRF~\cite{nerf}, NeAT~\cite{neat}, NAF~\cite{naf}, and TeRF (TensoRF~\cite{tensorf}) by \textbf{10.91},  \textbf{15.03},  \textbf{5.13}, and  \textbf{13.76 dB} in PSNR on the scenes of medicine, biology, security, and industry. The average gains are \textbf{over 12 dB}. The visual comparisons of our method and the second-best algorithms on four scenes (pelvis, bonsai, box, and engine) show that our method yields more perceptually pleasing results. 
	}
	\vspace{4mm}
	\label{fig:teaser}
}]

\vspace{4mm}
\begin{abstract}
    \vspace{-3mm}
    % Yuanhao Cai
    X-ray, known for its ability to reveal internal structures of objects, is expected to provide richer information for 3D reconstruction than visible light. Yet, existing NeRF algorithms overlook this nature of X-ray, leading to their limitations in capturing structural contents of imaged objects. In this paper, we propose a framework, Structure-Aware X-ray Neural Radiodensity Fields (SAX-NeRF), for sparse-view X-ray 3D reconstruction. Firstly, we design a Line Segment-based Transformer (Lineformer) as the backbone of SAX-NeRF. Linefomer captures internal structures of objects in 3D space by modeling the dependencies within each line segment of an X-ray. Secondly, we present a Masked Local-Global (MLG) ray sampling strategy to extract contextual and geometric information in 2D projection. Plus, we collect a larger-scale dataset X3D covering wider X-ray applications. Experiments on X3D show that SAX-NeRF surpasses previous NeRF-based methods by \textbf{12.56} and \textbf{2.49 dB} on novel view synthesis and CT reconstruction.  \url{https://github.com/caiyuanhao1998/SAX-NeRF}
\end{abstract}

\vspace{-1mm}
\let\thefootnote\relax\footnotetext{\text{\normalsize $\dag$ $=$  advisor,~~~~$*$ = corresponding authors}}

%%%%%%%%% BODY TEXT
\vspace{-5mm}
\section{Introduction}
\vspace{-1mm}
\label{sec:intro}

% 第一段定义任务，Background, definition, why they can be integrated into one framework. 一个问题是第一段话没把问题讲清楚。
Compared with natural light, X-ray has stronger penetrating power to reveal more internal structures of imaged objects. Hence, X-ray is widely used for prospective imaging~\cite{x_ray_1,x_ray_2,x_ray_3,x_ray_4} in  medicine, biology, security, industry, \emph{etc}. However, X-ray is harmful to human body because of its ionizing radiation. To reduce X-ray exposure, this paper studies the low-dose X-ray 3D reconstruction problem by decreasing X-ray imaging projections in the circular cone beam X-ray scanning scenario~\cite{cbct,mv_x_2,mv_x_1,mv_x_3,cbct_3}.
We focus on two tasks, \emph{i.e.}, novel view synthesis (NVS) and computed tomography (CT) reconstruction. NVS aims to create new projections of a scene from viewpoints not originally captured. CT reconstruction retrieves the 3D CT volume of the scanned object from multi-view X-ray projections. These two tasks are complementary with an overall objective to reconstruct 3D representations from 2D projections.

\begin{figure}[t]
	\centering
	\hspace{1mm}
	\includegraphics[width=\linewidth]{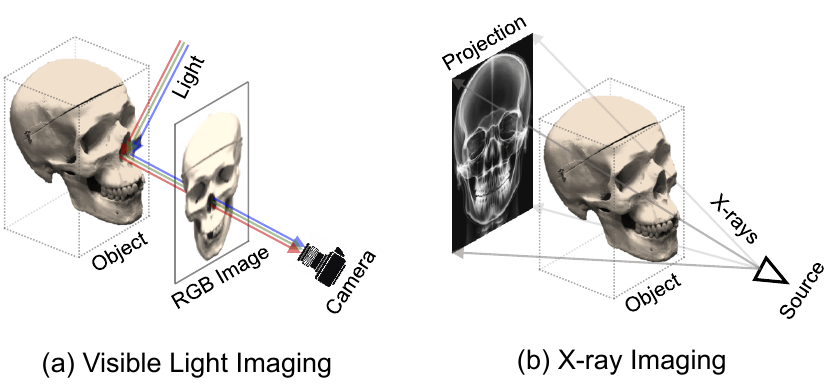} % 调整图片路径
	\vspace{-6mm}
	\caption{Visible light \emph{vs.} X-ray. Visible light imaging relies on reflection. X-ray imaging is based on penetration and attenuation.}
	\vspace{-4mm}
	\label{fig:imaging_compare}
\end{figure}

% 第二段要扣题，Structure-Aware, paired data 与 unpaired data
% 核心矛盾主要集中在NeRF-based这一块儿

A majority of existing deep learning-based methods employ a powerful model such as convolutional neural network (CNN) to learn a brute-force mapping from 2D  X-ray projections to 3D CT volumes. These methods require a large number of projection-CT pairs for training. Yet, CT volumes are not accessible in practice. Collecting even a small projection-CT dataset is tedious, labor-intensive, and harmful to health. Plus, these paired learning-based methods fail in generalizing from one application to another due to the large domain discrepancy between different CT datasets.

Recently, the emergence of NeRF~\cite{nerf}  provides a more reasonable solution to X-ray 3D reconstruction. Compared with paired learning-based algorithms, NeRF-based methods do not require CT volumes for training. Instead, they only need projections of just one scene. Although RGB NeRF algorithms have been well developed, directly applying them for X-ray scenes may achieve suboptimal results due to the fundamental differences between visible light and X-ray imaging. As compared in Fig.~\ref{fig:imaging_compare}, visible light imaging relies on the reflection off the surface of an object. It mainly captures external features. In contrast, X-rays penetrate the object and attenuate, thereby forming an image. X-ray imaging primarily reveals internal structures, which provide key clues for X-ray 3D reconstruction.

%   These methods need careful manual tweaking and usually require a long time. They also suffer from limited representation capacity and poor generalization ability. 

Nonetheless, current NeRF-based methods overlook this critical property of X-ray imaging. \textbf{Firstly}, they learn NeRF by a simple multilayer perceptron (MLP). X-ray attenuates differently when penetrating different structures. However, MLP treats each point on an X-ray equally, showing limitations in modeling 3D structures of objects. \textbf{Secondly}, previous methods mainly adopt a naive pixel-level ray sampling strategy in the training phase. They randomly sample X-rays corresponding to scattered pixels on the whole image coordinate system. As a result, the contextual information and geometric structures in 2D projection are not well extracted. Plus, X-ray projections are spatially sparse. Sampling X-rays on uninformative regions may lead to low efficiency. \textbf{Besides}, existing methods mainly study X-ray 3D reconstruction in limited medical scenes while their performance on other applications is still under-explored.

To tackle these issues, we propose a novel framework, Structure-Aware X-ray Neural Radiodensity Fields (SAX-NeRF), with the key insight of capturing 2D and 3D structures in X-ray imaging. \textbf{Firstly}, we design a Line Segment-based Transformer (Lineformer) as the backbone of SAX-NeRF. It partitions an X-ray into different line segments and then samples points on each one. By computing self-attention within every piece of the X-ray, Lineformer can model internal dependencies and learn complex 3D structures of different parts penetrated by the X-ray. Unlike vanilla Transformer~\cite{vaswani2017attention} whose computational cost is quadratic to the number of input points, Lineformer is more efficient by enjoying linear computational complexity. \textbf{Secondly}, we present a Masked Local-Global (MLG) ray sampling strategy. It uses a binary mask to segment informative foreground regions on the projection. We crop non-overlapping patches from these informative regions and then sample X-rays that land on the pixels inside these patches to help Lineformer perceive local contextual information and 2D structures. For the informative regions outside the patches, we randomly sample X-rays to help Lineformer perceive the scene's 2D global shape and geometry. \textbf{Besides}, we collect a larger-scale dataset, X3D, to evaluate the performance of X-ray 3D reconstruction algorithms in wider application scenarios. As shown in Fig.~\ref{fig:teaser}, our SAX-NeRF surpasses state-of-the-art (SOTA) NeRF-based methods by large margins on the NVS task. The average improvements on all scenes of X3D are \textbf{over 12 dB}.

Our contributions can be summarized as follows:
\begin{itemize}
	\vspace{0.5mm}
	%\item We propose a new unpaired learning-based framework, SAX-NeRF, for sparse-view X-ray 3D reconstruction.
        \item We propose a novel method, SAX-NeRF, for sparse-view X-ray 3D reconstruction without CT data for training.
	\vspace{0.5mm}
	\item We present a new Transformer, Lineformer, to capture complex internal structures of imaged objects in 3D space. To our knowledge, it is the first attempt to explore the potential of Transformer in X-ray neural rendering.
	\vspace{0.5mm}
	\item We design an MLG sampling strategy to extract geometric and contextual information of objects in 2D projection.
	\vspace{-3.3mm}
	\item We establish a larger-scale benchmark, X3D, for X-ray 3D reconstruction. Experiments show that our method outperforms SOTA methods on NVS and CT reconstruction tasks across different application scenarios of X-ray.
\end{itemize}

\vspace{-0.3mm}
\section{Related Work}
\label{sec:rela}
\vspace{-0.2mm}

\subsection{Neural Rendering}
\vspace{-0.2mm}

NeRF~\cite{nerf} represents objects via an implicit function of color and volume density, yielding high-quality results on the NVS task. 
Follow-up works improve NeRF with more fine-grained details \cite{mipnerf21, barron2022mipnerf360} and broader applications  \cite{chen2022hallucinated, wang2022rodin,suhail2022light}.
Meanwhile, to reduce the computational cost of NeRF, learnable feature encodings \cite{instant_ngp, tensorf, chen2023neurbf} are designed to embed input point positions.
% Later works \cite{naf, mednerf} further extend NeRF-based rendering into radiological images. 
However, applying existing RGB NeRF methods for X-ray rendering~\cite{naf,mednerf,neat} may achieve suboptimal results due to the differences between visible light and X-ray imaging. For instance, NAF~\cite{naf} follows NeRF to employ an MLP model for medical X-ray neural rendering, showing limitations in capturing complex structures of imaged objects in 3D space and 2D projection.

%To remedy this issue, researchers introduce NeRF-based algorithms~\cite{bibid}

\begin{figure*}[htp]
	\begin{center}
		\begin{tabular}[t]{c} \hspace{-2.8mm}			\includegraphics[width=1\textwidth]{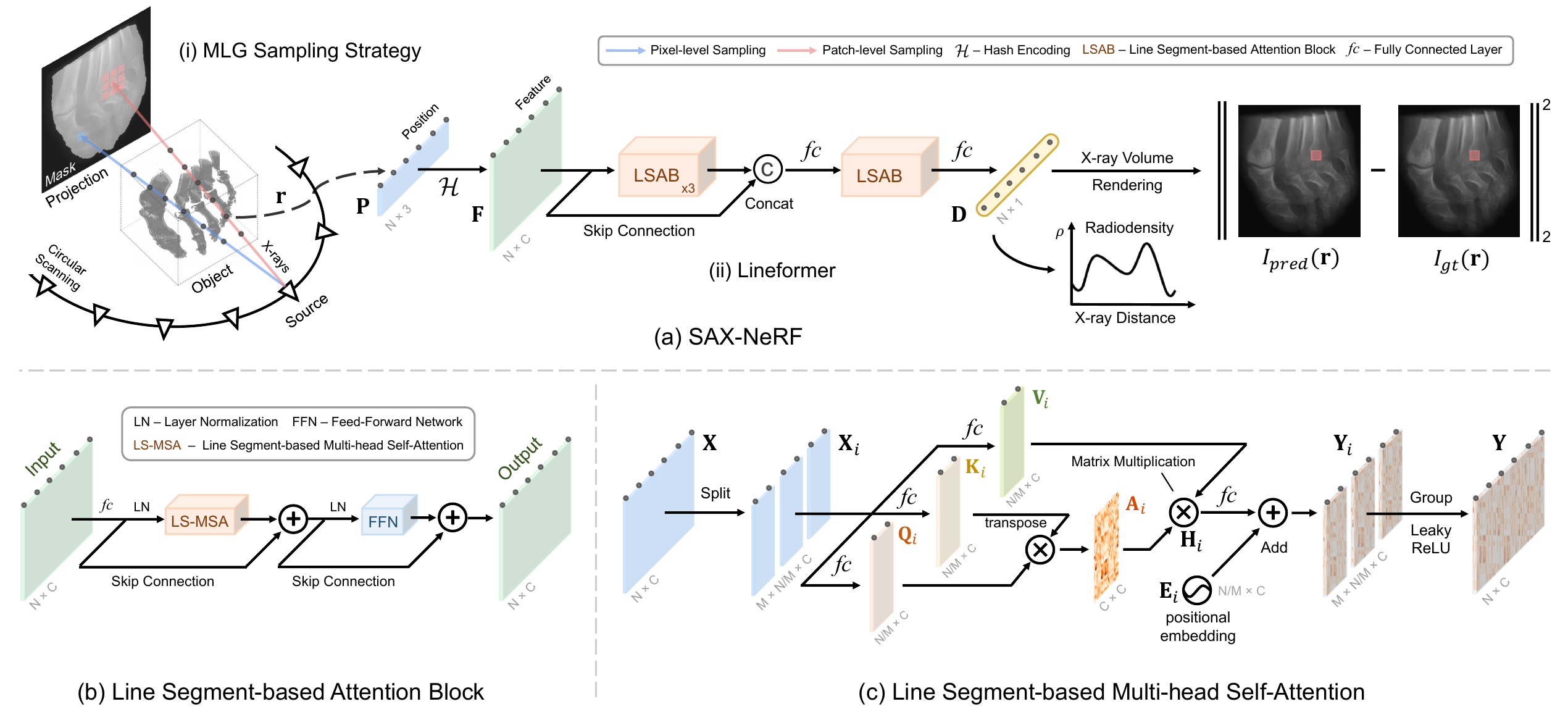}
		\end{tabular}
	\end{center}
	\vspace*{-6mm}
	\caption{\small Overview of our method. (a) SAX-NeRF uses (i) MLG strategy to sample an X-ray batch $\mathcal{R}$. Then $N$ point positions $\mathbf{P}$ on each X-ray $\mathbf{r} \in \mathcal{R}$ are sampled and input into (ii) Lineformer to produce the radiodensity $\mathbf{D}$. (b) Line Segment-based Attention Block (LSAB) is the basic unit of Lineformer. It captures inner structural dependencies by (c) Line Segment-based Multi-head Self-Attention (LS-MSA). }
	\label{fig:pipeline}
	\vspace{-2mm}
\end{figure*}

\subsection{Cone Beam CT Reconstruction}
Traditional cone beam CT Reconstruction algorithms are mainly divided into two categories: analytical methods~\cite{fdk,analytical_2} and optimization-based methods~\cite{asd_pocs,sart,vw_art,iterative_1,iterative_2,iterative_3}. Analytical methods predict the CT volume by solving the Radon transformation~\cite{radon} and its inverse. %For example, Feldkamp \emph{et al.}~\cite{fdk} propose FDK that accumulates intensities by backprojecting from 2D views. 
These methods can achieve good results when given hundreds of projections but fail in handling sparse-view cases. Optimization-based algorithms treat the reconstruction as a \emph{maximum a posteriori} (MAP) problem based on hand-crafted image priors and solve it by iteratively minimizing the energy function, which takes a long time. %For instance, Sidky \emph{et al.}~\cite{asd_pocs} propose  ASD-POCS that minimizes total variation of images. However, the optimization process takes a long time and the hand-crafted image prior limits the representing ability. 
Recently, CNNs~\cite{x2ct,dif_net,ctnet,ct_transfer,intratomo} and diffusion models~\cite{diffusion_mbir} have been applied to CT reconstruction and achieve good results.
%Ying~\emph{et al.}~\cite{x2ct} design X2CT-GAN to explore CT reconstruction from biplanar X-rays.  
Yet, these methods require a number of data pairs for training. 
%They also struggle with poor generalization ability due to the large domain discrepancy between different CT images. 
To avoid the above restrictions, we develop NeRF-based algorithms.

%\vspace{-1mm}
\subsection{Vision Transformer}
%\vspace{-1mm}
Transformer~\cite{vaswani2017attention} is first proposed for machine translation. In recent years, it has achieved great success in computer vision including image classification~\cite{xcit,arnab2021vivit,global_msa}, object detection~\cite{to_2,de_detr,DETR}, semantic segmentation~\cite{SETR,segformer,ts_1}, image restoration~\cite{restormer,retinexformer,dauhst,mst} and generation~\cite{styleswin,gat,transgan}, \emph{etc.} Nonetheless, directly applying vanilla Transformer for X-ray neural rendering will suffer from expensive computational cost with respect to the number of input points. The potential of Transformer for X-ray neural rendering still remains under-explored. We aim to fill this research gap.

%\vspace{-2mm}
\section{Method}
\label{sec:method}
%\vspace{-1mm}

%In this scenario, a scanner emits cone-shaped X-ray beams to capture sparse-view 2D projections at equal angular intervals, as shown in the left part of Fig.~\ref{fig:pipeline} (a).

\subsection{Overall Framework}
%\vspace{-1mm}
Fig.~\ref{fig:pipeline} illustrates the pipeline of our method. The left part of  Fig.~\ref{fig:pipeline} (a) depicts the scenario of circular cone beam X-ray scanning where a scanner emits cone-shaped X-ray beams and captures sparse-view projections at equal angular intervals. We first use (i) Masked Local-Global (MLG) strategy to sample a batch of X-rays  $\mathcal{R}$ landing on the projections for training. Then $N$ point positions $\mathbf{P} = \{\mathbf{p}_1, \cdots, \mathbf{p}_N\} \in \mathbb{R}^{N\times 3}$ are sampled on each X-ray $\mathbf{r} \in \mathcal{R}$ and fed into (ii) Lineformer. The basic unit of Lineformer is Line Segment-based  Attention Block (LSAB). As shown in Fig.~\ref{fig:pipeline} (b), an LSAB consists of a fully connected ($fc$) layer, two layer normalization (LN), a feed-forward network (FFN), and a Line Segment-based Multi-head Self-Attention (LS-MSA). The details of LS-MSA are depicted in Fig.~\ref{fig:pipeline} (c). 

Firstly, we review RGB NeRF. An MLP with weights $\Theta$ is usually employed to learn the mapping function $F_{\Theta}$ from the point position $(x, y, z) \in \mathbb{R}^3$ at the view direction $(\theta, \phi)$  to the color $(R, G, B) \in \mathbb{R}^3$ and volume density $\sigma \in \mathbb{R}$ as
\vspace{-0.7mm}
\begin{equation}
	\small
	F_{\Theta}: ~(x, y, z, \theta, \phi) \rightarrow (R, G, B, \sigma).
	\vspace{-0.7mm}
\end{equation}
As shown in Fig.~\ref{fig:imaging_compare}, visible light of specific wavelengths reflects off the surface of the object, thus revealing its color. In contrast, X-rays penetrate the object, thereby not reflecting the color information. Instead, it records the radiodensity property that denotes the degree to which a substance blocks or attenuates the passage of X-rays or other ionizing radiation. Since the radiodensity only depends on the point position, we aim to model the neural radiodensity fields as
\vspace{-1.1mm}
\begin{equation}
	\small
	F_{\Theta_L}: ~(x, y, z) \rightarrow  \rho,
	\vspace{-1.1mm}
	\label{eq:radiodensity_field}
\end{equation}
where $F_{\Theta_L}$ represents the mapping function of our Lineformer with weights $\Theta_L$ and $\rho \in \mathbb{R}$ denotes the radiodensity. According to the Beer-Lambert law, the intensity of an X-ray is reduced by the exponential integration of the traversed object's radiodensity. Hence, the ground-truth intensity $I_{gt}\mathbf{(r)} \in \mathbb{R}$  of the X-ray $\mathbf{r}(t) = \mathbf{o} + t \mathbf{d} \in \mathbb{R}^3$ with the near and far bounds $t_n$ and $t_f \in \mathbb{R}$ can be formulated as
\vspace{-0.7mm}
\begin{equation}
\small
I_{gt}\mathbf{(r)} = {I}_0 \cdot \text{exp}\big({-\int_{t_n}^{t_f}\rho(\mathbf{r}(t)) dt}\big),
\vspace{1.3mm}
\label{eq:intensity_render}
\end{equation}
where $I_0$ is the initial intensity. By discretizing Eq.~\eqref{eq:intensity_render}, we derive the predicted projection intensity $I_{pred}(\mathbf{r}) \in \mathbb{R}$ as
\vspace{-0.3mm}
\begin{equation}
	\small
	{I}_{pred}(\mathbf{r}) = {I}_0 \cdot \text{exp}\big({-\sum_{i=1}^{N} \rho_i \delta_i}\big),
	\label{eq:intensity_render_discrete}
	\vspace{-0.2mm}
\end{equation}
where $\rho_i$ denotes the predicted radiodensity of the $i$-th sampled point and $\delta_i = || \mathbf{p}_{i+1} - \mathbf{p}_{i} ||$ is the distance between adjacent points. Eventually, the training objective is to minimize the total squared error $\mathcal{L}$ between the predicted and ground-truth intensities in the training X-ray batch $\mathcal{R}$ as
\vspace{0.1mm}
\begin{equation}
\small
\mathcal{L} = \sum_{\mathbf{r} \in \mathcal{R}} \Big|\Big|~ {I}_{pred}(\mathbf{r}) - {I}_{gt}\mathbf{(r)} ~\Big|\Big|^2_2~,
\vspace{-0.1mm}
\end{equation}
where $I_{gt}(\mathbf{r})$ is obtained from the pixel value on  projection. $\mathcal{L}$ is depicted by the red pixels in the right part of Fig.~\ref{fig:pipeline} (a).

\subsection{Line Segment-based Transfomer}
As aforementioned, X-ray imaging reveals internal structures of imaged objects, which provide key clues for 3D reconstruction. Yet, previous methods overlook this important imaging property. Specifically, similar to RGB NeRF algorithms, existing X-ray NeRF methods~\cite{naf,intratomo} mainly adopt a simple MLP model to learn the implicit neural representations. X-ray attenuates differently when penetrating different structural contents. Yet, the MLP model treats each sampled point on an X-ray equally, showing limitations in modeling the 3D structures penetrated by the X-ray.

Towards this issue, we propose a Line Segment-based Transformer (Lineformer), as shown in Fig.~\ref{fig:pipeline} (a) (ii). The point position $\mathbf{P}$ is firstly fed into a hash encoding~\cite{instant_ngp} module $\mathcal{H}$ to produce point feature $\mathbf{F} \in \mathbb{R}^{N \times C}$ as $\mathbf{F} = \mathcal{H}(\mathbf{P})$. Then $\mathbf{F}$ undergoes four LSABs with a skip connection and two $fc$ layers to derive the point radiodensity $\mathbf{D} \in \mathbb{R}^{N}$.

LSAB is the basic unit of Lineformer. Its most important component is the LS-MSA mechanism, which captures internal structural dependencies by computing self-attention within each line segment of an X-ray. As illustrated in Fig.~\ref{fig:pipeline} (c), the input point feature $\mathbf{X} \in \mathbb{R}^{N \times C}$ is firstly partitioned into $M$ segments along the point dimension as
\vspace{-0.5mm}
\begin{equation}
	\small
	\mathbf{X} = [\mathbf{X}_1,~\mathbf{X}_2,~\cdots~,~ \mathbf{X}_M]^{\text{T}},
	\label{eq:partition}
	\vspace{-0.3mm}
\end{equation}
where $\mathbf{X}_i \in \mathbb{R}^{\frac{N}{M} \times C}$ and $ i = 1, 2, \cdots , M$. Then each $\mathbf{X}_i$ is linearly projected into  \emph{query} $\mathbf{Q}_i \in \mathbb{R}^{\frac{N}{M} \times C}$, \emph{key} $\mathbf{K}_i \in \mathbb{R}^{\frac{N}{M} \times C}$, and \emph{value} $\mathbf{V}_i \in \mathbb{R}^{\frac{N}{M} \times C}$ by three $fc$ layers as
\vspace{0.1mm}
\begin{equation}
	\small
	\mathbf{Q}_i = \mathbf{X}_i\mathbf{W}^{\mathbf{Q}_i},~~~\mathbf{K}_i = \mathbf{X}_i\mathbf{W}^{\mathbf{K}_i},~~~\mathbf{V}_i = \mathbf{X}_i\mathbf{W}^{\mathbf{V}_i},
	\label{linear_proj}
%	\vspace{0.3mm}
\end{equation}
where $\mathbf{W}^{\mathbf{Q}_i}$, $\mathbf{W}^{\mathbf{K}_i}$, and $\mathbf{W}^{\mathbf{V}_i} \in \mathbb{R}^{C \times C}$ are learnable parameters of the $fc$ layers; $biases$ are omitted for simplification. Subsequently, $\mathbf{W}^{\mathbf{Q}_i}$, $\mathbf{W}^{\mathbf{K}_i}$, and $\mathbf{W}^{\mathbf{V}_i}$ are uniformly split into $k$ heads along the channel dimension as
%\vspace{-0.5mm}
\begin{equation}
\small
\begin{aligned}
    \mathbf{Q}_i &= [\mathbf{Q}_i^1,~\mathbf{Q}_i^2,~\cdots~,~\mathbf{Q}_i^k], \\
    \mathbf{K}_i &= [\mathbf{K}_i^1,~\mathbf{K}_i^2,~\cdots~,~\mathbf{K}_i^k], \\
    \mathbf{V}_i &= [\mathbf{V}_i^1,~\mathbf{V}_i^2,~\cdots~,~\mathbf{V}_i^k].
\end{aligned}
%\vspace{-0.5mm}
\end{equation}
The dimension for each head is $d_h = C/k$. Fig.~\ref{fig:pipeline} (b) illustrates the situation with $k = 1$ for simplicity. Then the self-attention within each head $\mathbf{H}_i^j$ is computed as
\vspace{-0.6mm}
\begin{equation}
	\small
	\mathbf{H}_i^j = \text{Attn}(\mathbf{Q}_i^j, \mathbf{K}_i^j, \mathbf{V}_i^j) = \mathbf{V}_i^j ~\text{softmax}(\frac{{\mathbf{K}_i^j}^{\text{T}}\mathbf{Q}_i^j}{\alpha_i^j}),
	\label{eq:ls_msa}
	\vspace{-0.8mm}
\end{equation}
where $\alpha_i^j \in \mathbb{R}$ is a learnable parameter that adaptively scales the inner product before the softmax function. Successively, $k$ heads are concatenated in channel dimension to pass through an $fc$ layer and then plus a positional embedding $\mathbf{E}_i \in \mathbb{R}^{\frac{N}{M} \times C}$ to derive the $i$-th output $\mathbf{Y}_i \in \mathbb{R}^{\frac{N}{M} \times C}$ as
\vspace{-1.5mm}
\begin{equation}
	\small
	\mathbf{Y}_i = [\mathbf{H}_i^1,~\mathbf{H}_i^2,~\cdots~,~\mathbf{H}_i^k]~\mathbf{W}_i + \mathbf{E}_i,
	\vspace{1.5mm}
\end{equation}
where $\mathbf{W}_i \in \mathbb{R}^{C \times C}$ are learnable parameters of the $fc$ layer. Finally, we group the outputs of $M$ segments in point dimension to obtain the output feature $\mathbf{Y} \in \mathbb{R}^{{N} \times C}$ as
\vspace{-0.4mm}
\begin{equation}
	\small
	\mathbf{Y} = [\mathbf{Y}_1,~\mathbf{Y}_2,~\cdots~,~ \mathbf{Y}_M]^{\text{T}}.
	\vspace{-0.4mm}
\end{equation}
By capturing the interactions of points within each line segment, the proposed Lineformer is more capable of perceiving the complex internal 3D structures of different parts penetrated by the X-ray and therefore modeling the implicit neural radiodensity fields in Eq.~\eqref{eq:radiodensity_field} more accurately.

\vspace{2mm}
\noindent\textbf{Complexity Analysis.} We analyze the computational complexity of our LS-MSA and compare it with the global multi-head self-attention (G-MSA) mechanism of vanilla Transformer. The computational cost of LS-MSA primarily comes from the two matrix multiplication, \emph{i.e.}, $\mathbb{R}^{d_h \times \frac{N}{M}} \times \mathbb{R}^{\frac{N}{M} \times d_h}$ and $\mathbb{R}^{\frac{N}{M} \times d_h} \times \mathbb{R}^{d_h\times d_h}$, in Eq.~\eqref{eq:ls_msa} performed $k \times M$ times. Thus, the complexity of LS-MSA is formulated as
\vspace{0.4mm}
\begin{equation}
	\small
	\begin{aligned}
		\mathcal{O}(\text{LS-MSA}) &= kM\cdot[d_h\cdot(d_h\cdot \frac{N}{M}) + \frac{N}{M}\cdot(d_h\cdot d_h)], \\
		&= 2kMd_h^2\frac{N}{M} = 2Nk(\frac{C}{k})^2 = \frac{2NC^2}{k}.
	\end{aligned}
	\label{complexity_lsmsa}
	\vspace{1.2mm}
\end{equation}
While the complexity of G-MSA is formulated as
%\vspace{-0.5mm}
\begin{equation}
	\small
	\mathcal{O}(\text{G-MSA}) = 2N^2C.
	\label{complexity_gmsa}
	%\vspace{-0.5mm}
\end{equation}
Compare Eq.~\eqref{complexity_lsmsa} with Eq.~\eqref{complexity_gmsa}. $\mathcal{O}(\text{G-MSA})$ is quadratic to the number of input points ($N$). This heavy computational burden impedes the application of Transformer for X-ray 3D reconstruction. In contrast, $\mathcal{O}(\text{LS-MSA})$ is linear to $N$. This significantly reduced cost allows for the integration of LS-MSA into each basic unit LSAB of Lineformer, thereby further exploring the tremendous potential of Transformer.

%\vspace{-0.4mm}
\subsection{Masked Local-Global Ray Sampling}
\begin{figure}[t]
	\centering
	\hspace{-2.6mm}
	\includegraphics[width=1.01\linewidth]{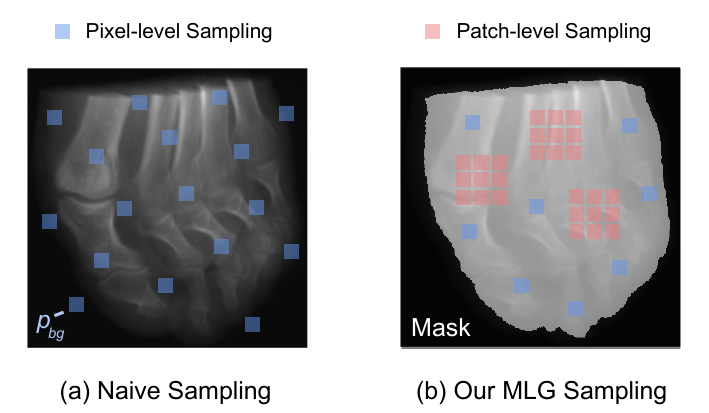} % 调整图片路径
	\vspace{-2mm}
	\caption{Comparison of ray sampling. (a) The naive strategy samples X-rays that land on scattered pixels. (b) Our MLG strategy performs pixel- and patch-level sampling on foreground regions. }
	\vspace{-2mm}
	\label{fig:mlg}
\end{figure}

%\vspace{-0.4mm}
As shown in Fig.~\ref{fig:mlg} (a), existing NeRF algorithms mainly adopt a naive pixel-level ray sampling strategy. They randomly sample X-rays corresponding to scattered pixels on the whole image coordinate system for training. This naive strategy has two drawbacks. \textbf{Firstly}, it shows limitations in extracting local contextual and geometric representations in 2D projection because the semantic information from neighbor pixels is not captured. \textbf{Secondly}, X-ray images are spatially sparse. Some randomly sampled X-rays may land on the background dark regions of the projection, such as the pixel $p_{bg}$ in Fig.~\ref{fig:mlg} (a). These X-rays do not penetrate the object and thus are not imaged on the projection. In other words, these X-rays are uninformative because they do not characterize the radiodensity property of the object. Learning with these X-rays will degrade the model efficiency.

To address these problems, we propose a Masked Local-Global (MLG) ray sampling strategy, as shown in Fig.~\ref{fig:mlg} (b). MLG first uses a mask $\mathbf{M} \in \mathbb{R}^{H\times W}$ to segment the imaged foreground regions. $\mathbf{M}$ is derived by binarizing the projection $\mathbf{I} \in \mathbb{R}^{H\times W}$ with a threshold $T \in \mathbb{R}$ as $\mathbf{M} = \mathbf{1}_{\mathbf{I}>T}$. Subsequently, to avoid redundant sampling, we partition $\mathbf{M}$ into a set $\mathcal{W} \in \mathbb{R}^{\frac{HW}{S^2} \times S\times S}$ of $\frac{HW}{S^2}$ non-overlapping windows with size $S\times S$. Let $\mathcal{W}_f$ denote the set of windows that are entirely contained in the foreground regions as
%\vspace{-0.9mm}
\begin{equation}
	\small
	\mathcal{W}_f = \{\mathbf{w} \in \mathcal{W}~|~\mathbf{w} = \mathbf{1}_{S \times S}\}.
	\label{eq:patch_sampling}
	%\vspace{-0.9mm}
\end{equation}
To capture local semantic information of the object, we perform patch-level sampling. Specifically, we randomly select $N_l$ windows $\mathcal{W}_l = \{\mathbf{w}_1, \cdots, \mathbf{w}_{N_l}\}$ from $\mathcal{W}_f$, as shown in the red patches of Fig.~\ref{fig:mlg} (b). Then the X-ray set $\mathcal{R}_l$ corresponding to the pixels within $\mathcal{W}_l$ can be formulated as
\vspace{-1mm}
\begin{equation}
	\small
	\mathcal{R}_l = \bigcup_{i = 1}^{N_l}\bigcup_{p \in \mathbf{w}_i} \text{Ray}(p),
	\vspace{-1mm}
\end{equation}
where $\text{Ray}(p)$ is a function that maps from a pixel $p$ to its corresponding X-ray. Furthermore, to assist the model in better capturing global contextual representations and perceiving the overall geometric shape of the imaged object, we perform pixel-level sampling. Particularly, we randomly select $N_g$ pixels $\mathcal{P}$ from the foreground regions excluding the area of $\mathcal{W}_l$ to avoid repeated ray sampling, as depicted in the blue pixels of Fig.~\ref{fig:mlg} (b). $\mathcal{P}$ can be formulated as
\vspace{-0.0mm}
\begin{equation}
	\small
	\mathcal{P} = \{p \in (\mathbf{M} - \mathcal{W}_l)~|~p = 1\}.
	\vspace{-0.4mm}
\end{equation}
Then the X-ray set $\mathcal{R}_g$ corresponding to  $\mathcal{P}$ is obtained by
\vspace{-1mm}
\begin{equation}
	\small
	\mathcal{R}_g = \bigcup_{p \in \mathcal{P}} \text{Ray}(p).
	\vspace{-1mm}
\end{equation}
Finally, the training X-ray batch is the union of $\mathcal{R}_l$ and $\mathcal{R}_g$:
\vspace{-2.3mm}
\begin{equation}
	\small
	\mathcal{R} = \mathcal{R}_l \bigcup \mathcal{R}_g.
	\vspace{1.3mm}
\end{equation}
Using MLG ray sampling strategy, the model can more effectively capture the contextual information and model the geometric structures of the imaged object on 2D projection. %In the testing phase, X-rays according to all pixels on the whole image coordinate system are sampled and inferred.
%and extract the rays  corresponding to the pixels of the selected windows

\begin{table*}[t]
	\centering
	\setlength\tabcolsep{4pt}
	\resizebox{1\textwidth}{!}{\hspace{-0.5mm}
		%\begin{tabular}{l|cc|cc|cc|cc|cc|cc|cc|cc}
		\begin{tabular}{l|cccccccccc|cc}
			\toprule[0.15em]
			%\multirow{2}{*}{Methods}      & \multicolumn{10}{c}{Medical}   & \multicolumn{3}{c}{Biology} & Security & Industry & Total    \\  
			Method &\multicolumn{2}{c}{~~~~InTomo~\cite{intratomo}~~~~~~} & \multicolumn{2}{c}{~~~~~~~~NeRF~\cite{nerf}~~~~~~~~~~~~~}  &\multicolumn{2}{c}{~~~~~~~~NeAT~\cite{neat}~~~~~~~~~~~}
			&\multicolumn{2}{c}{~~~~~~~TensoRF~\cite{tensorf}~~~~~~~~~~~} &\multicolumn{2}{c|}{~~~~~~~~NAF~\cite{naf}~~~~~~~~~~} &\multicolumn{2}{c}{~~\textbf{SAX-NeRF}~~~~} \\
			Scene~~~~~~~~ &~~PSNR~~ &~~SSIM~~ &~~PSNR~~ &~~SSIM~~ &~~PSNR~~ &~~SSIM~~ &~~PSNR~~ &~~SSIM~~ &~~PSNR~~ &~~SSIM~~ &~~PSNR~~ &~~SSIM~~ \\ \midrule[0.15em]
			%Method &PSNR &SSIM &PSNR &SSIM &PSNR &SSIM &PSNR &SSIM&PSNR &SSIM &PSNR &SSIM \\ \midrule[0.15em]

			Jaw &26.91 &0.9937 &34.84 &0.9981 &32.68 &0.9911  &34.06  &0.9964 &\ul{39.89} &\ul{0.9988} &\bf 42.75 &\bf 0.9992\\
			
			Leg &42.53 &0.9976 &45.92 &\ul{0.9989} &47.71 &0.9981  &41.40 &0.9969 &\ul{50.87} &0.9988 &\bf 56.86 &\bf 0.9996\\
			
			Box &34.65 &0.9963 &35.67 &\ul{0.9985} &\ul{36.14} &0.9957  &35.43 &0.9977 &35.98 &0.9955 &\bf 39.67 &\bf 0.9992\\
			
			Carp &24.04 &0.9648 &20.62 &0.9467 &31.26 &0.9620  &\ul{37.35}  &\ul{0.9973} &29.60 &0.9593 &\bf 59.88 &\bf 0.9999\\
			
			Foot &39.48 &0.9979 &\ul{41.05} &\ul{0.9989} &38.24 &0.9963  &37.73 &0.9929 &38.35 &0.9913 &\bf 46.64 &\bf 0.9994\\
			
			Head  &\ul{34.83}  &0.9977 &29.76 &\ul{0.9991} &27.74 &0.9295 &34.43 &0.9878  &30.17 &0.9531 &\bf 53.06 &\bf 0.9995\\
			
			Pelvis &38.72 &0.9961 &40.79 &0.9972 &37.70 &0.9866 &41.57 &0.9948  &\ul{43.76} &\ul{0.9975} &\bf 53.27 &\bf 0.9995\\
			
			Chest  &28.95 &0.9915 &36.16  &0.9988 &40.77 &0.9990  &23.61 &0.9402  &\ul{42.37} &\ul{0.9993} &\bf 47.42 &\bf 0.9994\\
			
			Bonsai &39.26 &0.9953 &37.67  &0.9983 &47.02 &0.9985  &47.80 &\ul{0.9989}  &\ul{49.03} &\ul{0.9989} &\bf 55.33 &\bf 0.9995\\
			
			Teapot &41.51 &0.9978 &34.66 &\ul{0.9993} &29.29 &0.9669  &\ul{44.18} &\ul{0.9993}  &34.92 &0.9985 &\bf 52.62 &\bf 0.9996\\
			
			Engine &23.99 &0.9517 &21.07 &0.9334 &30.36  &0.8854  &\ul{39.72} &\ul{0.9918} &31.68 &0.9195 &\bf 58.80 &\bf 0.9998\\
			
			Pancreas  &20.03 &0.8537 &19.85 &0.8560 &\ul{37.53} &\ul{0.9017}  &29.24 &0.8031  &36.23 &0.8844 &\bf 49.88 &\bf 0.9978\\
			
			Abdomen    &27.64  &0.9646 &24.62 &0.9559 &26.74 &0.8563  &27.38 &0.8730  &\ul{37.59} &\ul{0.9855} &\bf 54.22 &\bf 0.9996\\
			
			Aneurism     &20.81 &0.9621 &24.97 &0.9792 &35.41 &0.9936  &\ul{47.99} &\ul{0.9997}  &39.62 &0.9990 &\bf 52.91 &\bf 0.9998\\
			
			Backpack &36.09  &0.9918 &39.75 &0.9962 &41.60 &0.9969  &\ul{43.16} &0.9977 &42.02  &\ul{0.9982} &\bf 47.17 &\bf 0.9989\\
			
			\midrule[0.15em]
			
			Average &31.96 &0.9768 &32.49 &0.9770 &36.01  &0.9638  &37.67 &0.9712  &\ul{38.81} &\ul{0.9785} &\bf 51.37 &\bf 0.9994\\
			
			\bottomrule[0.15em]
	\end{tabular}}
	\vspace{-1mm}
	\caption{Quantitative comparisons on the novel view synthesis task. The best results are in \textbf{bold} and the second-best results are \ul{underlined}. }\label{tab:quantitative_nvs}
	\vspace{-1mm}
\end{table*}

\begin{figure*}[t]
	\begin{center}
		\begin{tabular}[t]{c} \hspace{-3.6mm}
			\includegraphics[width=1\textwidth]{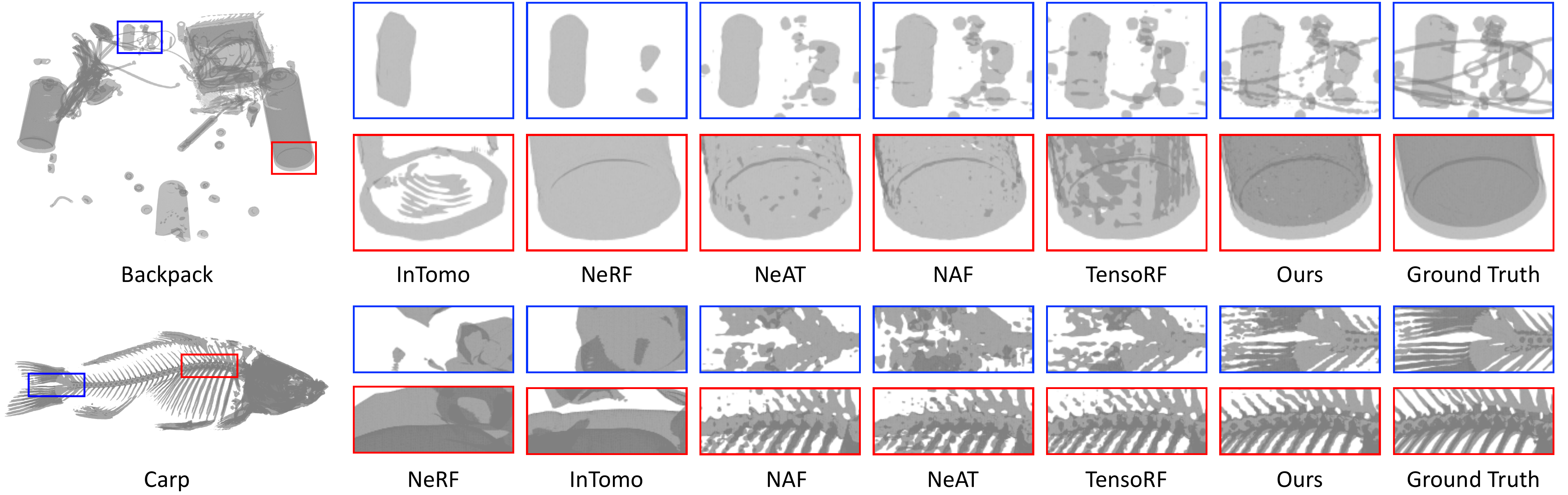}
		\end{tabular}
	\end{center}
	\vspace*{-6mm}
	\caption{\small Qualitative results of novel view synthesis on the scenes of backpack (top) and carp (bottom). Please zoom in for a better view.}
	\label{fig:proj_vis_compare}
	\vspace{-2mm}
\end{figure*}

\begin{table*}[t]
	\centering
	\setlength\tabcolsep{4pt}
	\resizebox{1\textwidth}{!}{\hspace{-0.8mm}
		%\begin{tabular}{l|cc|cc|cc|cc|cc|cc|cc|cc}
		\begin{tabular}{l|cccccccccccccccc|cc}
			\toprule[0.15em]
			%\multirow{2}{*}{Methods}      & \multicolumn{10}{c}{Medical}   & \multicolumn{3}{c}{Biology} & Security & Industry & Total    \\  
			Method&\multicolumn{2}{c}{FDK~\cite{fdk}} &\multicolumn{2}{c}{ASD-POCS~\cite{asd_pocs}} &\multicolumn{2}{c}{SART~\cite{sart}} &\multicolumn{2}{c}{InTomo~\cite{intratomo}}~~ & \multicolumn{2}{c}{NeRF~\cite{nerf}}  &\multicolumn{2}{c}{NeAT~\cite{neat}}
			&\multicolumn{2}{c}{TensoRF~\cite{tensorf}}
			&\multicolumn{2}{c|}{NAF~\cite{naf}} &\multicolumn{2}{c}{\textbf{SAX-NeRF}} \\
			Scene~~~~ &PSNR &SSIM &PSNR &SSIM &PSNR &SSIM &PSNR &SSIM &PSNR &SSIM &PSNR &SSIM &PSNR &SSIM &PSNR &SSIM &PSNR &SSIM \\ \midrule[0.15em]
			%Method &PSNR &SSIM &PSNR &SSIM &PSNR &SSIM &PSNR &SSIM&PSNR &SSIM &PSNR &SSIM \\ \midrule[0.15em]

			Jaw &28.58 &0.7816 &33.25 &0.9325 &33.13 &0.9301 &31.95 &0.9162 &32.17 &0.9114 &32.53 &0.9139 &31.90 &0.8971 &\ul{34.14} &\ul{0.9358} &\bf 35.47  &\bf 0.9525\\
			
			Leg &28.48 &0.6690 &35.39 &0.9826 &35.30 &0.9809 &36.41 &0.9882 &39.27 &0.9938 &40.29 &0.9902 &40.70 &0.9923 &\ul{41.28} &\ul{0.9940} &\bf 43.47 &\bf 0.9973\\
			
			Box &24.14 &0.5616 &31.27 &0.9226 &31.20 &0.9200 &30.59 &0.9140 &\ul{33.58} &\ul{0.9494} &31.58 &0.9298 &32.17 &0.9314  &31.78 &0.9309 &\bf 35.33 &\bf 0.9602\\
			
			Carp &32.32 &0.8177 &37.63 &\ul{0.9777} &36.89 &0.9682 &32.47 &0.9493 &32.99 &0.9529 &36.85 &0.9576 &37.52 &0.9687  &\ul{37.93} &0.9711 &\bf 42.72 &\bf 0.9902\\
			
			Foot &24.53 &0.6000 &29.98 &0.9208 &30.29 &0.9296 &31.43 &0.9127 &30.03  &0.9072 &30.86 &0.9221 &30.46 &0.9153 &\ul{31.63} &\ul{0.9363} &\bf 32.25 &\bf 0.9403\\
			
			Head  &26.17 &0.7155 &35.27 &0.9707 &34.88 &0.9597 &31.07 &0.9303 &34.15 &0.9672 &35.56 &0.9679 &35.53 &0.9672  &\ul{36.46} &\ul{0.9743} &\bf 39.70 &\bf 0.9888\\
			
			Pelvis &26.91 &0.6367 &34.26 &0.9493 &34.38 &0.9481 &30.38 &0.9042 &31.72 &0.9170 &33.73 &0.9370 &35.13 &0.9528  &\ul{36.01} &\ul{0.9654} &\bf 40.40 &\bf 0.9870\\
			
			Chest  &22.89 &0.7861 &31.13 &0.9422 &32.17 &\ul{0.9594} &22.04 &0.7460 &28.40 &0.8925 &31.20 &0.9497 &30.13 &0.9308  &\ul{33.05} &0.9581 &\bf 34.38 &\bf 0.9718\\
			
			Bonsai &24.53 &0.7276 &32.70 &0.9529 &33.02 &\ul{0.9600} &28.90 &0.8811 &31.77 &0.9382 &33.20 &0.9476 &33.47 &0.9521 &\ul{33.85} &0.9585 &\bf 36.51 &\bf 0.9761\\
			
			Teapot &31.07 &0.8059 &37.35 &0.9800 &37.38 &0.9787 &36.15 &0.9786 &41.67 &\ul{0.9945} &40.85 &0.9872 &\ul{42.71} &0.9942 &42.56  &0.9926 &\bf 44.32 &\bf 0.9970\\
			
			Engine &23.02 &0.5405 &30.81 &0.9580 &30.44 &0.9442 &27.49 &0.9264 &36.85 &0.9858 &36.63 &0.9804 &35.21 &0.9728 &\ul{37.84} &\ul{0.9859} &\bf 38.77 &\bf 0.9917\\
			
			Pancreas  &9.641 &0.1232 &18.30 &0.7701 &18.36 &0.7008 &16.01 &0.7865 &17.73 &\ul{0.8614} &19.06 &0.8541 &\ul{19.75} &0.7737 &19.41 &0.8126 &\bf 22.98 &\bf 0.9531\\
			
			Abdomen    &22.63 &0.6030 &31.46 &0.9231 &31.40 &0.9170 &28.05 &0.8754 &29.71 &0.9049 &31.14 &0.9060 &31.51 &0.9073 &\ul{34.45} &\ul{0.9501} &\bf 35.01 &\bf 0.9598\\
			
			Aneurism     &28.07 &0.7295 &34.73 &0.9864 &34.76 &0.9864 &30.32 &0.9652 &31.97 &0.9353 &35.80 &0.9819 &37.36 &\ul{0.9889} &\ul{37.73} &0.9871 &\bf 41.46 &\bf 0.9956\\
			
			Backpack &23.84 &0.5351 &31.34 &0.9309 &31.32 &0.9294 &28.77 &0.8753 &30.28 &0.9192 &31.90 &0.9345 &33.16 &0.9362 &\ul{33.26} &\ul{0.9501} &\bf 35.97 &\bf 0.9688\\
			
			\midrule[0.15em]
			
			Average &25.12 &0.6422 &32.32 &0.9400 &32.33 &0.9342 &30.29 &0.9189 &32.15 &0.9354 &33.41 &0.9447 &33.78 &0.9387 &\ul{34.76} &\ul{0.9535} &\bf 37.25 &\bf 0.9753\\
			
			\bottomrule[0.15em]
	\end{tabular}}
	%\vspace{1.5mm}
	\vspace{-2mm}
	\caption{Quantitative comparisons on the CT reconstruction task. The best results are in \textbf{bold} and the second-best results are \ul{underlined}.  }\label{tab:quantitative_ct}
	\vspace{-1mm}
\end{table*}

\begin{figure*}[t]
	\begin{center}
		\begin{tabular}[t]{c} \hspace{-2.8mm}
			\includegraphics[width=1.0\textwidth]{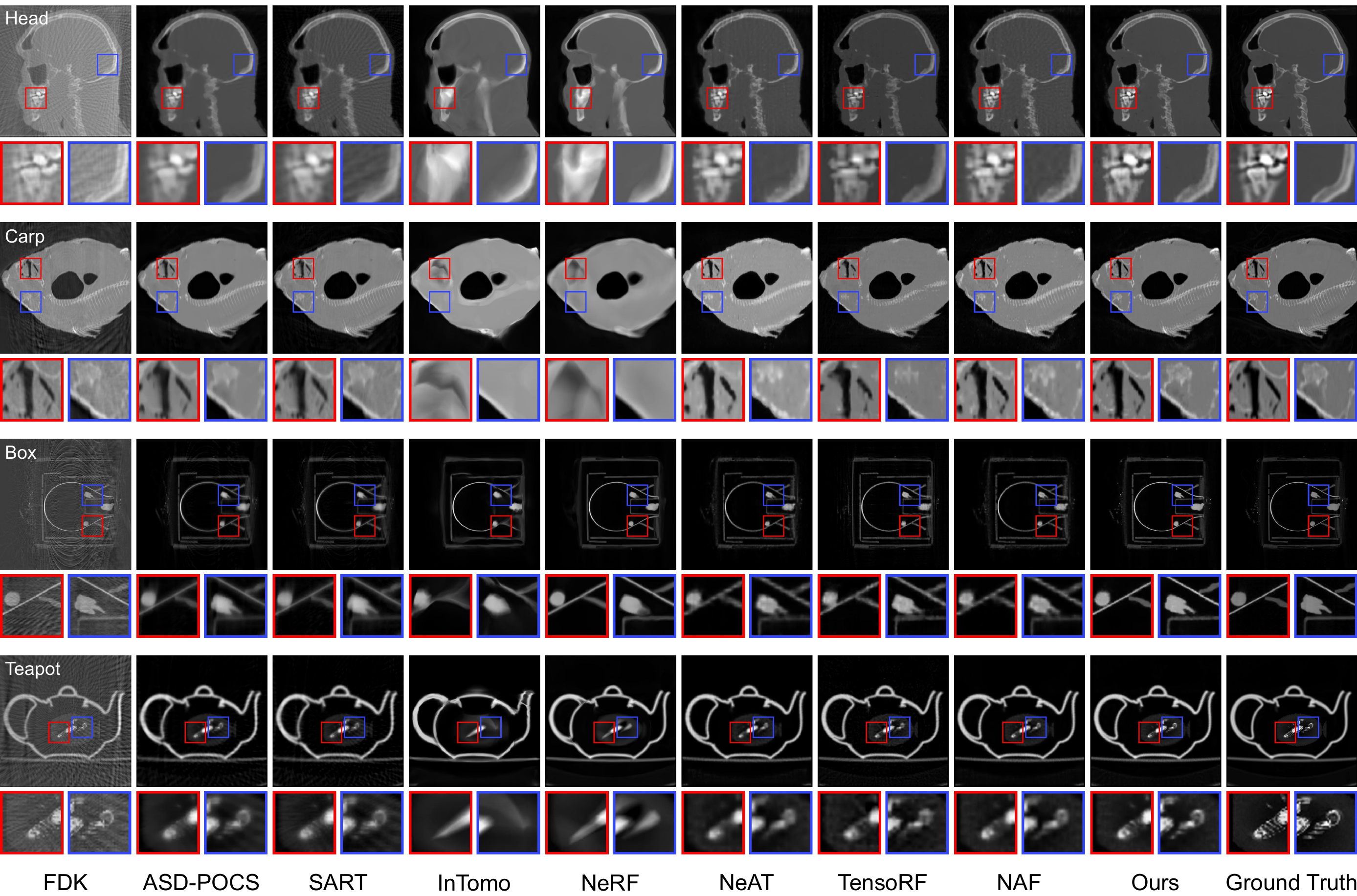}
		\end{tabular}
	\end{center}
	\vspace{-5mm}
	\caption{\small Visual results of CT reconstruction on the scenes of head, carp, box, and teapot (top to bottom). Please zoom in for a better view.}
	\label{fig:ct_vis_compare}
	\vspace{-1mm}
\end{figure*}

%\vspace{-1mm}
\section{Experiment}
\label{sec:exp}
%\vspace{-0.5mm}

\subsection{Experimental Settings}
\vspace{0mm}

\noindent\textbf{X3D Dataset.} Previous methods mainly conduct X-ray 3D reconstruction research on limited medical applications. For example, NAF~\cite{naf} is evaluated on five medical scenes. The performance of NeRF-based methods on other X-ray applications is under-explored. To fill this research gap, we collect a larger-scale dataset, X3D, containing 15 scenes and covering 4 applications, \emph{i.e.}, medicine, biology, security, and industry. We collect the CT volumes of X3D from public datasets. Specifically, the chest, backpack, carp, and pancreas datasets are collected from LIDC-IDRI~\cite{LIDC-IDRI}, MIDA~\cite{backpack}, D$^2$VR~\cite{carp}, and DeepOrgan~\cite{pancreas}, respectively. The teapot, aneurism, bonsai, and foot datasets are obtained from VOLVIS~\cite{volvis} and the rest are from the open scientific visualization dataset~\cite{osvd}. Then we use the tomographic method TIGRE~\cite{tigre} to generate projections by scanning CT volumes with 3\% noise in the range of 0$^{\circ}$ $\sim$ 180$^{\circ}$.

\vspace{1mm}
\noindent\textbf{Implementation Details.} We implement our SAX-NeRF by PyTorch~\cite{pytorch}. The model is trained with the Adam~\cite{adam} optimizer ($\beta_1$ = 0.9 and $\beta_2$ = 0.999) for 3000 iterations. The learning rate is initially set to 1$\times$10$^{-4}$ and is halved every 1500 iterations during the training procedure. The batch size of X-rays is set to 2048, 1024 of which is from patch-level sampling and the other 1024 is from pixel-level sampling. We uniformly sample 320 points along each X-ray. For each scene, we use its 50 projections to train, another 50 projections to test the performance of NVS, and its CT volume to evaluate the results of CT reconstruction. All experiments are conducted on an RTX 8000 GPU. We adopt the peak signal-to-noise ratio (PSNR) and structural similarity index measure (SSIM)~\cite{ssim} as the evaluation metrics.

%将四个表格合并成一个
\begin{table*}[t]\hspace{-0.5mm}
	% Table  a - break down ablation
	\subfloat[\small Break-down ablation to higher performance.\label{tab:breakdown}]{\vspace{1mm}
		\scalebox{0.64}{
			\begin{tabular}{c c c c c}
				%\small
				\toprule
				\rowcolor{color3} Baseline &LS-MSA &MLG   &NVS &CT \\
				\midrule
				\checkmark & & &37.97 / 0.9748 &34.21 / 0.9513 \\
				\checkmark  &\checkmark & &47.95 / 0.9945 &36.86 / 0.9717 \\
				\checkmark & &\checkmark &43.51 / 0.9861 &35.30 / 0.9601 \\
				\checkmark  &\checkmark &\checkmark &\bf 51.37 / 0.9994 &\bf 37.25 / 0.9753 \\
				\bottomrule
	\end{tabular}}}\hspace{1mm}\vspace{0mm}
	% Table c - study of MSA
	\subfloat[\small Analysis of the line segment quantity. \label{tab:line_segments}]{
		\vspace{1mm}\scalebox{0.64}{
			\begin{tabular}{l c c c c c}
				\toprule
				\rowcolor{color3} Num &20 &40 &80 &160 &320\\
				\midrule
				\multirow{2}{*}{NVS}  &45.776 &47.832 &50.148 &\bf 51.365 &50.620 \\
				&0.9892 &0.9939 &0.9991 &\bf 0.9994 &0.9992  \\
				\multirow{2}{*}{CT}  &35.845 &36.787 &37.088 &\bf 37.249 &37.186 \\
				&0.9670 &0.9711 &0.9745 &\bf 0.9753 &0.9749 \\
				\bottomrule
	\end{tabular}}}\vspace{0mm}
	% Table b - study of ORF
	\subfloat[\small Analysis of the sampling patch size.\label{tab:patch_size}]{\vspace{1mm}
		\scalebox{0.64}{
			\begin{tabular}{l c c c c c}
				\toprule
				\rowcolor{color3} Size &~2$\times$2~ &~4$\times$4~ &~8$\times$8~ &~16$\times$16~ &~32$\times$32~\\
				\midrule
				\multirow{2}{*}{NVS}  &48.515 &\bf 51.365 &50.297 &50.408 &49.629 \\
				&0.9976 &\bf 0.9994 &0.9992 &0.9992 &0.9989  \\
				\multirow{2}{*}{CT}  &36.982 &\bf 37.249 &37.118 &37.163 &37.014 \\
				&0.9733 &\bf 0.9753 &0.9748 &0.9749 &0.9741 \\
				\bottomrule
	\end{tabular}}}\hspace{1mm}\vspace{0mm}
	\vspace{-2mm}
	\caption{\small We conduct ablation study on all scenes of X3D. Average PSNR and SSIM are reported on the NVS and CT reconstruction tasks.}
	\label{tab:ablations}\vspace{-1mm}
\end{table*}

\begin{figure*}[t]
	\centering
	\hspace{-3.6mm}
	\begin{minipage}{0.57\textwidth}
		\centering
		\includegraphics[width=\linewidth]{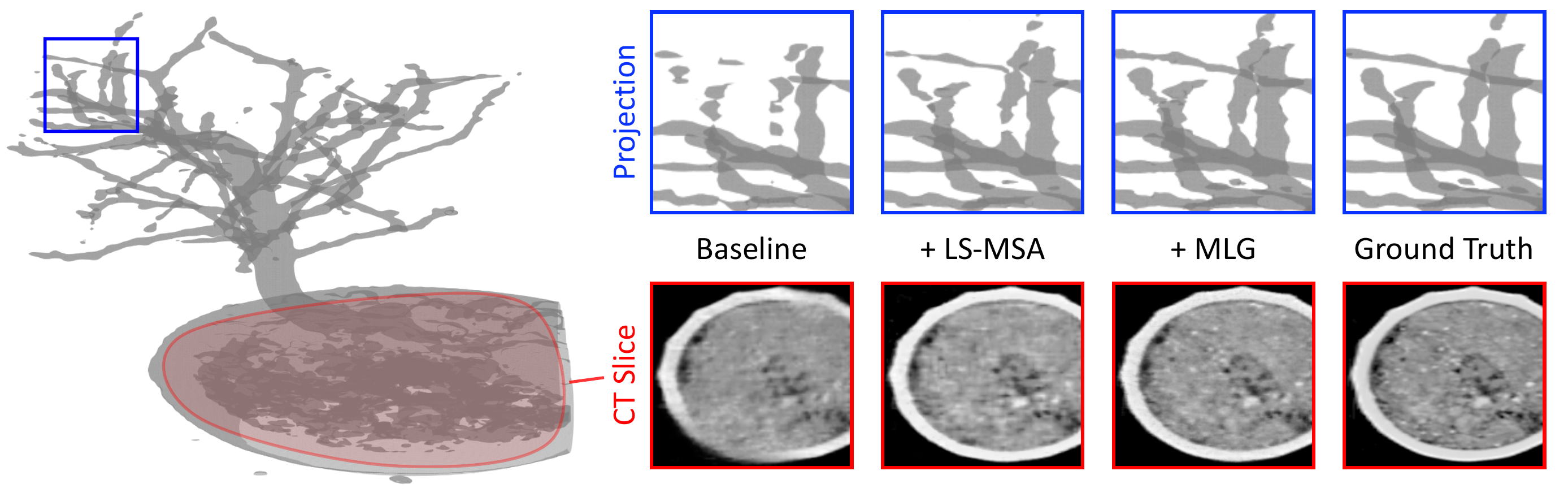} % 调整图片路径
		\caption{Visual analysis. Using LS-MSA and MLG  captures more structures.}
		\label{fig:visual_analysis}
	\end{minipage}\hspace{0.5mm}
	\begin{minipage}{0.43\textwidth}
		\centering
		\includegraphics[width=\linewidth]{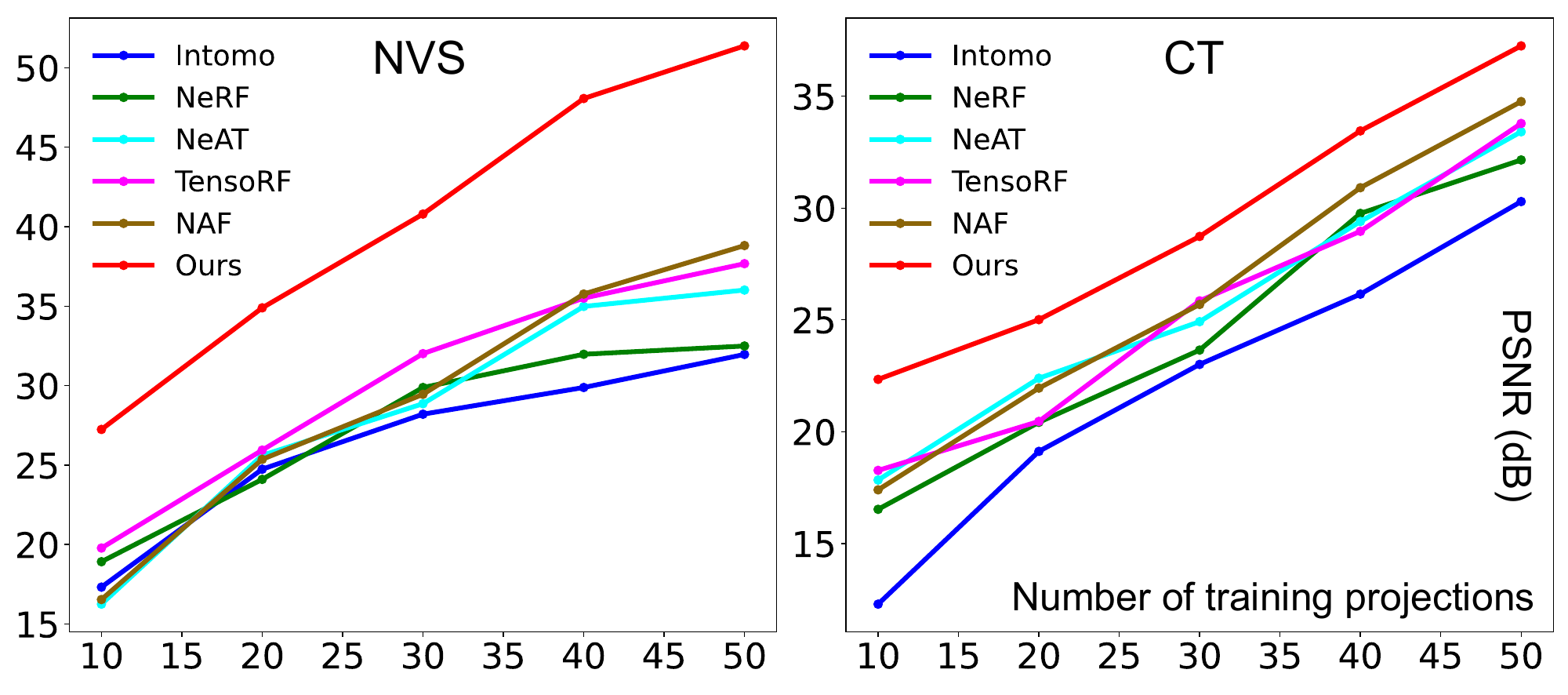} % 调整图片路径
		\vspace{-5.5mm}
		\caption{Analysis of the number of training projections.}
		\label{fig:proj_num}
	\end{minipage}
	\vspace{-2.5mm}
\end{figure*}

\subsection{Main Results}

\noindent\textbf{Novel View Synthesis.} Tab.~\ref{tab:quantitative_nvs} lists the quantitative results of PSNR and SSIM on the NVS task.  
We compare our SAX-NeRF with five SOTA NeRF-based algorithms including InTomo~\cite{intratomo}, NeRF~\cite{nerf}, NeAT~\cite{neat}, TensoRF~\cite{tensorf}, and NAF~\cite{naf}. The input and output of all methods are set the same as Eq.~\eqref{eq:radiodensity_field} for fair comparison. It can be observed that our SAX-NeRF significantly outperforms SOTA methods on all scenes. Specifically, when compared with the recent best general RGB NeRF algorithm TensoRF, our SAX-NeRF is 13.70 dB (51.37 - 37.67) and 0.0282 (0.9994 - 0.9712) higher in PSNR and SSIM. When compared with the recent best medical NeRF method NAF, SAX-NeRF surpasses it by 12.56 dB in PSNR and 0.0209 in SSIM. The average improvements of our method on the scenes of medicine, biology, security, and industry are 10.91, 15.03, 5.13, and 13.76 dB, as shown in the bar charts of Fig.~\ref{fig:teaser}. 

The qualitative results are depicted in Fig.~\ref{fig:proj_vis_compare}. As can be observed from the zoomed-in patches, previous methods are less effective in synthesizing novel projections. They either produce blurry images or fail to reconstruct structural contents. In contrast, our SAX-NeRF yields more visually pleasing results with clearer textures and more fine-grained details while preserving more complete geometric structures. More visual comparisons are shown in Fig.~\ref{fig:teaser}.

% 讲清楚四类分别包括哪些。
\vspace{1mm}
\noindent\textbf{CT Reconstruction.} Tab.~\ref{tab:quantitative_ct} reports the quantitative results on the CT reconstruction task. For fairness, we do not compare projection-CT paired learning-based algorithms, but instead focus on comparing methods that only require X-ray projections of single scenes for training or direct processing. In addition to the five SOTA NeRF-based algorithms. We also compare SAX-NeRF with an analytical method (FDK~\cite{fdk}) and two optimization-based algorithms (ASD-POCS~\cite{asd_pocs} and SART~\cite{sart}). Our method yields the best results on all scenes. In particular, SAX-NeRF dramatically outperforms previous NeRF-based, optimization-based, and analytical algorithms by over 2.49, 4.92, and 12.13 dB.

Fig.~\ref{fig:ct_vis_compare} displays the visual comparisons in four application scenarios including medicine (head), biology (carp), security (box), and industry (teapot). Other methods either produce over-smooth images blurring the structural contents or introduce distracting artifacts. In contrast, our SAX-NeRF is more favorable to reconstruct vivid high-frequency details such as sharp edges while maintaining spatial smoothness of homogeneous regions within complex structures.

These results convincingly demonstrate the advantages of the proposed SAX-NeRF in X-ray 3D reconstruction.

%\vspace{-0.5mm}
\subsection{Ablation Study}
%\vspace{-0.5mm}

To reliably evaluate the effectiveness of our approaches, we conduct ablation study on all scenes of X3D and report the average PSNR / SSIM results in the following part.

\vspace{0.5mm}
\noindent\textbf{Break-down Ablation.} We adopt a baseline model that is derived by directly removing the LS-MSA module and MLG sampling strategy from our SAX-NeRF to conduct the break-down ablation study. The results are listed in Tab.~\ref{tab:breakdown}. The baseline model yields 37.97 and 34.21 dB on NVS and CT reconstruction. When using LS-MSA, the baseline model gains by 9.98 and 2.65 dB. When we apply MLG sampling, the model achieves 5.54 and 1.09 dB improvements. When jointly exploiting the two techniques, the model is improved by 13.40 and 3.04 dB on the NVS and CT reconstruction tasks. This evidence clearly exhibits the efficacy of Lineformer and MLG sampling strategy.

\vspace{0.5mm}
\noindent\textbf{Visual Analysis.} To intuitively show the effectiveness of the two proposed approaches, we further conduct visual analysis on the scene of bonsai. As depicted in Fig.~\ref{fig:visual_analysis}, the baseline model fails to preserve the geometry like the tree branches in the projection and blurs high-frequency details such as the edges of the basin in the CT slice. When successively using LS-MSA and MLG sampling, the model reconstructs more structural contents and fine-grained textures. 

\vspace{0.5mm}
\noindent\textbf{Parameter Analysis.} We conduct parameter analysis regarding the number of line segments $M$ in Eq.~\eqref{eq:partition} and the patch size $S$ in Eq.~\eqref{eq:patch_sampling}. Please note that we keep the total number of sampled points and X-rays unchanged for fair comparison. When analyzing one parameter, we fix the other at its optimal value. The results are reported in Tab.~\ref{tab:line_segments} and \ref{tab:patch_size}. It is clear that, when using different $M$ and $S$, our SAX-NeRF stably outperforms the baseline model by over 7.81 and 1.64 dB on NVS and CT reconstruction. This evidence suggests the reliability of our method. The model's performance yields its maximum when $M$ = 160 and $S$ = 4.

\vspace{0.5mm}
\noindent\textbf{Robustness Analysis.} We conduct robustness analysis regarding the number of training projections to compare the performance of different methods when given fewer X-ray projection views. The results are plotted as two line charts in Fig.~\ref{fig:proj_num}, where the vertical axis is PSNR (in dB performance) and the horizontal axis is the number of training projections.  Our SAX-NeRF reliably surpasses SOTA methods by large margins when given different numbers of training projections on both NVS (left) and CT reconstruction (right) tasks. Surprisingly, when using even only 60\% of training projections, SAX-NeRF still outperforms other algorithms on the NVS task. These results clearly exhibit the superiority and robustness of our proposed method.

\vspace{0.5mm}
\noindent\textbf{Lineformer \emph{vs.} vanilla Transformer.} We replace  LS-MSA with G-MSA to conduct comparative experiments. The experimental results show that our Lineformer significantly outperforms vanilla Transformer by 5.30 and 1.28 dB on the NVS and CT reconstruction tasks while only requiring 3.41\% of vanilla Transformer's computational complexity.

\vspace{-2.5mm}
\section{Conclusion}
\vspace{-0.5mm}
\label{sec:conclusion}
In this paper, we focus on studying a core problem in sparse-view X-ray 3D reconstruction, \emph{i.e.}, how to effectively capture the various and complex structures penetrated by X-rays. To this end, we propose a novel framework SAX-NeRF. To model 3D structural dependencies in space, we design a Transformer, Lineformer, as the backbone of SAX-NeRF. Lineformer partitions an X-ray into different line segments and then computes self-attention within each piece of the X-ray. In addition, to extract 2D geometry and contextual representations in projection, we present an MLG ray sampling strategy that contains pixel- and patch-level sampling on the informative foreground regions. Besides, we also collect a larger-scale dataset, X3D, covering wider X-ray application scenarios. Comprehensive experiments on X3D show that SAX-NeRF significantly surpasses SOTA algorithms on the NVS and CT reconstruction tasks.

\vspace{1mm}
\noindent \textbf{Acknowledgements:} This work was supported by the Lustgarten Foundation for Pancreatic Cancer Research and the Patrick J. McGovern Foundation Award.

{
    \small
    \bibliographystyle{ieeenat_fullname}
    \bibliography{main}
}


\end{document}